\documentclass[lnbip]{svmultln}
\usepackage{amsmath,amssymb,amsfonts}%
\usepackage{hyperref}
\usepackage{makeidx}  
\begin{document}
\mainmatter              
\title{Advanced Detection of Source Code Clones via an Ensemble of Unsupervised Similarity Measures}
\titlerunning{Advanced Detection of Source Code Clones}  
%
\author{Jorge Martinez-Gil}
\authorrunning{Jorge Martinez-Gil}   

\institute{Software Competence Center Hagenberg GmbH \\ Softwarepark 32a, 4232 Hagenberg, Austria \\ \url{jorge.martinez-gil@scch.at}}

\maketitle              

\begin{abstract}        
The capability of accurately determining code similarity is crucial in many tasks related to software development. For example, it might be essential to identify code duplicates for performing software maintenance. This research introduces a novel ensemble learning approach for code similarity assessment, combining the strengths of multiple unsupervised similarity measures. The key idea is that the strengths of a diverse set of similarity measures can complement each other and mitigate individual weaknesses, leading to improved performance. Preliminary results show that while Transformers-based CodeBERT and its variant GraphCodeBERT are undoubtedly the best option in the presence of abundant training data, in the case of specific small datasets (up to 500 samples), our ensemble achieves similar results, without prejudice to the interpretability of the resulting solution, and with a much lower associated carbon footprint due to training. The source code of this novel approach can be downloaded from \url{https://github.com/jorge-martinez-gil/ensemble-codesim}.

\end{abstract}

\section{Introduction}
Automatically assessing the degree of similarity between source code fragments presents several challenges ranging from syntactic variability or the use of complex algorithmic structures to the employment of diverse abstraction levels across programming languages \cite{juergens2009code}. While established techniques and tools exist to address this problem, significant limitations must be addressed. For example, the best-performing approaches need vast training data, their insights are practically intelligible to the human operator, and their associated carbon footprint during the training phase does not bring any positive societal outcome \cite{key-martinez-jfis}.

In order to overcome these issues, this research proposes a novel ensemble approach for improved code similarity assessment. The method combines the outputs of multiple similarity measures to produce better similarity scores. This approach is based on the principle that the strengths of diverse similarity measures can complement each other and mitigate individual weaknesses, leading to improved systemic performance.

Extensive empirical evaluations on several datasets have demonstrated the effectiveness of our ensemble learning method. Our method outperformed traditional single-model approaches in all cases (both in scenarios with small and large datasets). Furthermore, our method achieved comparable performance to state-of-the-art methods when solving small datasets without reliance on deep learning techniques, making it a solution to consider under the circumstances such as the absence of significant amounts of training data, the need for interpretability, and the wish for a reduced carbon footprint when training.

Therefore, our contribution is not only to guide the aggregation of appropriate unsupervised similarity measures for code similarity but also to pave the way for applications such as source clone detection \cite{saini2018code} or code plagiarism assessment \cite{novak2019source} in realistic scenarios where there is not too much data for training. In this way, the major contributions of this study are:

\begin{itemize}
	\item We build upon previous work \cite{key-martinez-swqd} to curate a comprehensive collection of unsupervised similarity measures to compare the likeness of different code fragments. This collection facilitates a systematic exploration of the most effective measures for assessing source code similarity at low computational cost.
	\item Furthermore, we investigate the learning of ensembles through bagging and boosting techniques. We aim to develop models with enhanced code similarity assessment capabilities using the abovementioned similarity measures.
	\item We evaluate the proposed ensembles focuses on metrics such as precision, recall, and f-measure. Our findings show that ensembles outperform individual similarity measures and have potential to rival even the most sophisticated state-of-the-art techniques in the absence of significant training data.
\end{itemize}

The remainder of this paper is structured as follows: Section 2 provides the background on the code similarity assessment challenge and establishes its importance in the software development field. Section 3 explains the unsupervised similarity measures to address this challenge and how we can rely on them to build ensembles with superior performance. Section 4 empirically evaluates the ensembles introduced before using popular benchmark datasets. Section 5 offers a detailed analysis and discussion of the experimental results. Finally, the paper concludes with lessons learned and outlines future lines of work.

\section{State-of-the-art}
The existence of many programming languages and individual coding styles poses a challenge for determining the degree of likeness between source code fragments \cite{ain2019systematic}. Despite that, unsupervised similarity measures demonstrate reasonable adaptability when working with code written in multiple languages or according to various styles, so it can be accepted that they are good indicators of code similarity \cite{key-martinez-swqd}. This feature is also important in many scenarios, such as maintenance, whereby increased complexity is commonplace \cite{higo2002software}.

The research community widely acknowledges that the emerging strategies based on language models offer superior performance in this context \cite{karmakar2021pre}. However, our hypothesis is that unsupervised methods for code similarity evaluation offer several advantages for addressing key aspects that should not be overlooked:

\begin{itemize}
 	\item Capturing the meaning of code is a complex task, as functionally equivalent code fragments may exhibit superficial differences or, on the contrary, share surface-level similarities while having entirely different purposes. Some unsupervised techniques specifically focus on the semantic intent rather than syntactic details.
	\item Since unsupervised techniques eliminate the need for labeled training data, they reduce the overhead of establishing a ground truth. While labeled examples remain valuable for validating the effectiveness of these methods, these examples are not a prerequisite for deployment.
	\item Codebases often contain comments and other non-functional elements embedded within the core logic. Specific unsupervised methods can distinguish between meaningful code patterns and just incidental material.
	\item Unsupervised methods can accommodate the diversity of programming languages and coding styles. This mitigates the need for a complex strategy that attempts to account for all possible variations.
\end{itemize}

We acknowledge the superior capabilities of language models when confronted with this challenge. However, we also recognize and aim to demonstrate that alternative models, which excel in specific areas, could be employed under certain plausible conditions across a wide range of scenarios. For instance, these models may be more suitable in situations where only limited datasets are available or where the working logic of the proposed solution needs to be clearly understood.

\subsection{Unsupervised Code Similarity}
There are various techniques to assess the similarity between source code fragments, each with its unique perspective based on certain features of the code fragments being compared. In previous work \cite{key-martinez-swqd}, we perform a categorization of exiting approaches consisting of twenty-one distinct unsupervised strategies. Nonetheless, according to the same research study \cite{key-martinez-swqd}, it is unrealistic to think that these techniques alone can deliver performance equivalent to the state-of-the-art. 

\subsection{Supervised Code Similarity}
Supervised code similarity is a well-studied field and has yielded the best results to date \cite{wang2020detecting,wei2017supervised,white2016deep,zhang2019novel,yu2019neural}. Although among the existing approaches, CodeBERT \cite{key-codebert} and its variant GraphCodeBERT \cite{guo2020graphcodebert} stand out. CodeBERT is a variant of BERT \cite{key-Bert}, and empirical results show that it is the leading technology to understand source code expressed in different general-purpose programming languages. It excels at tasks like generating documentation, creating code, and comparing similar pieces of code. The idea behind CodeBERT is to combine techniques from language processing with an understanding of code and training on a wide range of programming languages and their associated comments. Results show CodeBERT is very good at detecting similarities between source code fragments, as demonstrated by its performance in the resolution of benchmark datasets for clone detection \cite{key-codebert}. However, an additional fine-tuning phase on the pre-trained model over the actual data is usually required to obtain optimal results.

\subsection{Contribution over the state-of-the-art}
This research introduces a novel ensemble learning approach for code similarity assessment. The rationale behind combining multiple unsupervised similarity measures is to benefit from the strengths of individual approaches while mitigating their weaknesses. This strategy is helpful in scenarios where deep learning is not optimal due to the absence of significant training data, when prioritizing model interpretability without significantly compromising performance, or even when seeking to minimize the environmental impact.

Our work's focus is eminently practical. In fact, we evaluate our ensembles on benchmark datasets to confirm the approach's superiority over traditional techniques. Furthermore, we offer our code as open-source since this availability might promote transparency and collaborative advancement.

\section{Ensemble Approach}
We propose using an ensemble of unsupervised similarity measures to improve the performance when assessing code similarity. The rationale behind combining several unsupervised similarity measures is to offer a more accurate evaluation than strategies using just one measure \cite{key-martinez-ijseke}. Therefore, we first define the problem, introduce the concept of an ensemble, and discuss bagging and boosting as the two most prominent techniques for building ensembles. These techniques for building ensembles help aggregate different unsupervised similarity measures, leading usually to an enhanced strategy.

\subsection{Problem Statement}
We can define code similarity as follows: Given a set of code fragments \(S = \{C_1, C_2, \ldots, C_n\}\), we aim to find a function \(f: S \times S \rightarrow [0, 1]\) that measures the similarity between any two fragments \(C_i\) and \(C_j\).

The function \(f\) should map a pair \((C_i, C_j)\) to a continuous value within the real interval \([0, 1]\):

\begin{enumerate}
	\item  \(f(C_i, C_j) = 0\) indicates absolute code dissimilarity.
	\item  \(f(C_i, C_j) = 1\) indicates equivalent code.
	\item  The value of \(f(C_i, C_j)\) increases with increasing similarity between \(C_i\) and \(C_j\).
\end{enumerate}

Some applications can be built from this definition \cite{roy2009comparison}. For example, code clone detection can be performed by applying a threshold value to differentiate between clones and non-clones \cite{karnalim2021explanation}. 

\[
    \text{Clone}(C_i, C_j) = 
    \begin{cases}
        \text{True}, & \text{if } f(C_i, C_j) > \theta \\
        \text{False}, & \text{otherwise}
    \end{cases}
\]

In addition, something similar can also be done to determine whether or not the code was plagiarized in an academic environment \cite{aniceto2021source}, for example. These two challenges of a very similar nature will be studied in more detail later in the paper.  

\subsection{Ensemble Definition}
Let \( S = \{s_1, s_2, \ldots, s_n\} \) be a set of code fragments and \( M = \{m_1, m_2, \ldots, m_{21}\} \) be a set of unsupervised similarity measures. An ensemble \( E \) aims to aggregate similarity measures such as:

\[ E(s_a, s_b) = \alpha_1 \cdot m_1(s_a, s_b) + \ldots + \alpha_{21} \cdot m_{21}(s_a, s_b) \]

where \(\alpha_i\) represents the weight of each similarity measure, and  \( \sum_{i=1}^{21} \alpha_i = 1 \).

Each \( m_i(s_a, s_b) \) returns a similarity score in the real range [0, 1], where 0 indicates no similarity, and 1 indicates equivalent code fragments. The ensemble \( E \), therefore, also returns a score in the real range [0, 1], effectively aggregating the individual similarity assessments into a measure.

There are several techniques for learning ensembles, almost all based on a purely data-driven strategy. In the context of this work, and for practical reasons, we focus on two: bagging and boosting. Other techniques, such as stacking, will not be considered here because they have already been addressed in similar problems in the past, such as \cite{key-martinez-mlwa}. Although an in-depth study of each of these techniques is beyond the scope of this study, we offer below a brief description of how these ensemble-building strategies operate.

\subsection{Bagging}
The idea behind bagging is to train multiple instances of the same model with different data subsets and aggregate their predictions \cite{breiman1996bagging}. 

Formally, let \( D \) denote the original dataset and \( M \) the number of subsets to create. The subsets \( D_i \) for \( i \in \{1, 2, \ldots, M\} \) are generated through random sampling with replacement. For each subset \( D_i \), a model \( f_i \) is trained. Given an unsupervised similarity measure \( S \), we compute the similarity score for each subset:

\[
S_i = S(f_i, D_i) \quad \text{for } i \in \{1, 2, \ldots, M\}.
\]

The final aggregated prediction is obtained by averaging the individual predictions from all models:

\[
\hat{y} = \frac{1}{M} \sum_{i=1}^M f_i(D).
\]

This technique helps reduce variance and improve the stability of the model.

\subsection{Boosting}
The idea behind boosting is to train multiple models, with each model instance focusing on the errors of the previous one \cite{bentejac2021comparative}. 

Formally, let \( D \) denote the dataset and \( N \) the number of boosting rounds. We start with an initial model \( f_1 \) and iterate sequentially through each round. For each round \( t \in \{1, 2, \ldots, N\} \), we train a model \( f_t \) with a focus on the errors made by the previous models. Let \( S \) denote a set of unsupervised similarity measures.

At each round, we update a weight distribution \( w_t \) over the dataset based on the errors of \( f_t \), aiming to emphasize the misclassified cases. The similarity scores for the current model \( f_t \) are calculated as:

\[
S_t = S(f_t, D) \quad \text{for } t \in \{1, 2, \ldots, N\}.
\]

The final aggregated prediction is obtained by combining the models using their respective weights:

\[
\hat{y} = \sum_{t=1}^N \alpha_t f_t(D),
\]

where \( \alpha_t \) is the weight assigned to each model \( f_t \) based on its performance in reducing the error. This sequential training allows for iterative refinement of the predictions.

\subsection{Differences with the Transformer architecture}
Using an ensemble of unsupervised semantic similarity measures for code similarity assessment offers several advantages over relying solely on a model like CodeBERT \cite{key-Bert}, or its more powerful variant GraphCodeBERT \cite{guo2020graphcodebert}. The ensemble approach combines different unsupervised similarity measures, each capturing unique aspects of code, thus providing a diverse analysis that can reduce individual biases and improve performance. 

Ensembles are also more flexible and scalable and do not require massive labeled data, making them suitable for diverse programming languages and code domains. Therefore, our approach benefits from the strengths of multiple unsupervised similarity measures, potentially outperforming all of them. Transformers-inspired solutions can face challenges with interpretability due to their complex attention mechanisms and might struggle with out-of-distribution or noisy code fragments. However, if sufficient data is available and there are no requirements beyond performance, results from transformer architectures should be much superior.

\section{Empirical Evaluation}
In previous work \cite{key-martinez-swqd}, we show that unsupervised similarity measures exhibit strengths in different domains. Some excel in textual comparisons, making them ideal for identifying cloned text. Others prioritize the detection of functional similarity, while specific measures specialize in uncovering structural resemblances between code and textual elements. The optimal choice of measure ultimately depends on the composition of the benchmark dataset. However, it is not reasonable to ask a human expert to do that since it would require many hours of work, and the result would be prone to errors since the task is far from trivial. The good news is that bagging and boosting algorithms, with their data-driven strategies, will decide the inclusion and importance of each of the simple unsupervised similarity measures.

\subsection{Baseline}
In this study, we compare our ensembles using two strategies. First, a weak baseline consists of obtaining better results than each similarity measure considered individually. Secondly, a strong baseline consists of the state-of-the-art conformed by CodeBERT \cite{key-Bert} and its variant GraphCodeBERT \cite{guo2020graphcodebert}. 

In this regard, the state-of-the-art models can have a previous fine-tuning phase that begins with loading and randomly splitting a dataset of code fragments into training, validation, and test sets. Then, the approach's core involves training the model to discern clone pairs, guided by a trainer configured with specific training arguments like epoch count, batch size, and learning rate adjustments. Finally, performance metrics are calculated to evaluate the model's effectiveness as the average value of a given number of executions. In our study, we considered up to 10 independent executions.

\subsection{Small-scale Dataset}
Firstly, we use the IR-Plag dataset \cite{key-karnalim}. This dataset was designed for benchmarking code similarity techniques. It contains code files deliberately crafted to simulate patterns of academic plagiarism. Although primarily intended for plagiarism detection, its characteristics align with our goals due to the overlap between plagiarized and cloned code, which involves replication (though the intent may differ). The IR-Plag dataset encompasses a variety of complexities without strictly classifying the clone types.

The dataset consists of seven original code files. A substantial portion, 355 files (77\%), are labeled as plagiarized, indicating a high prevalence of duplication. One hundred five files are considered non-plagiarized, potentially representing modified or derivative works. This yields a total of 467 code files. Collectively, these files contain 59,201 tokens, with 540 unique tokens underlining the dataset's lexical diversity. File sizes exhibit variation, ranging from 286 tokens (largest) to 40 tokens (smallest), with an average file size of approximately 126 tokens. Therefore, although small in size, it is a dataset with a high compositional diversity of programming elements.

\subsection{Large-scale Dataset}
Secondly, we use the BigCloneBench dataset, which is a benchmark collection created to evaluate how well different code clone detection strategies perform. It is helpful to tackle the challenge of finding duplicate code. It includes a variety of programming languages, and its data comes from actual software and student projects. This mix makes it an interesting resource for evaluating clone detection in real-life situations. This dataset is usually used in training solutions to make code more maintainable.

The dataset is divided into three subsets: training, validation, and testing. The training set contains 901,028 items, while the validation and testing sets contain 415,416 items each. This distribution is outlined to ensure comprehensive training, validation, and testing of the model. This helps compare different strategies and determine which ones are best at finding code clones.

\subsection{Evaluation Criteria}
In principle, measuring the accuracy of the different techniques is appropriate. That is the percentage of occasions in which the strategy under study succeeded or failed in predicting the similarity of code fragments. However, this metric is strongly discouraged for unbalanced datasets because the mere fact of constantly repeating the label of the most numerous classes in the test set will yield a high value that may be misleading.

For that reason, the community prefers other metrics to assess the fitness a clone detection strategy. For instance, the Precision/Recall evaluation method is popular because it emphasizes the importance of positive classes by separately assessing false positives (precision) and false negatives (recall). It penalizes the model for not identifying positive cases and for incorrect positive predictions. Subsequently, one can compute a harmonic mean of these two metrics, known as the F-measure, which serves as a basis for an homogeneous ranking.

\subsection{Hardware requeriments}
Please note that all the computations of similarity measures and the learning of their ensembles have been done on a CPU 11th Gen Intel(R) Core(TM) i7-1185G7 a 3.00GHz with 32GB for RAM (TDP of 50W approx.), while the configuration and execution of the baselines with CodeBERT and GraphCodeBERT has required a GPU Tesla V100-PCIE-16GB (TDP of 300W approx.).

\subsection{Results}
Table \ref{table:ss} shows us the results obtained for the first benchmark dataset, i.e., the small-scale IR-Plag. It can be seen that some unsupervised similarity measures are pretty good on their own if one looks at the f-measure. However, it is striking that ensembles obtain the best results, so we far exceed our weak baseline. Please note that we do not study time consumption issues here as they have been studied previously \cite{key-martinez-swqd}.

\begin{table}
\centering
\begin{tabular}{|l|c|c|c|}
\hline
\textbf{Approach} & \textbf{Precision} & \textbf{Recall} & \textbf{F-Measure} \\ \hline
Abstract Syntax Tree &  0.95 & 0.30 & 0.45 \\ 
Bag-of-Words &  0.93 & 0.19 & 0.31 \\ 
CodeBERT &  0.91 & 0.03 & 0.05 \\ 
Comment Sim. &  0.88 & 0.40 & 0.55 \\ 
Output Analysis &  0.88 & 0.93 & 0.90 \\ 
Function Calls &  0.85 & 0.67 & 0.75 \\ 
Fuzzy Matching &  0.82 & 0.50 & 0.62 \\ 
Graph Matching &  0.82 & 0.91 & 0.86 \\ 
Rolling Hash & 0.94 & 0.20 & 0.33 \\ 
Perceptual Hash &  0.89 & 0.41 & 0.57 \\ 
Jaccard &  0.99 & 0.26 & 0.42 \\ 
Longest Common Subsequence &  0.78 & 0.97 & 0.87 \\ 
Levenshtein &  0.82 & 0.76 & 0.79 \\ 
Metrics comparison &  0.81 & 0.57 & 0.67 \\ 
N-Grams &  0.84 & 0.83 & 0.84 \\ 
Program Dependence Graph &  0.85 & 0.39 & 0.53 \\ 
Rabin-Karp &  0.87 & 0.47 & 0.61 \\ 
Semantic Clone & 0.94 & 0.24 & 0.38 \\
Semdiff method & 0.87 & 0.25 & 0.39 \\ 
TDF-IDF &  0.97 & 0.17 & 0.29 \\ 
Winnow & 0.89 & 0.65 & 0.75 \\ \hline
\textbf{Boosting} & 0.88 & 0.99 & 0.93 \\
\textbf{Bagging} & 0.95 & 0.97 & 0.96 \\
\hline
\end{tabular}
\caption{Summary of the results obtained for the IR-Plag dataset}
\label{table:ss}
\end{table}

Table \ref{table:ss2} establishes a comparative for the state-of-the-art published to this dataset. On the one hand, it can be seen that the ensembles can perform better than CodeBERT. The reason is that they perform better without large data for training. This is even though several combinations of different partitions have been evaluated: 60-20-20, 70-15-15, and 80-10-10 on training, validation, and test, respectively. We are reporting here the best average obtained. On the other hand, GraphCodeBERT matches our best results. However, with the disadvantages already discussed throughout this study, it might make sense to use our solution when working with small datasets.

\begin{table}
\centering
\begin{tabular}{|l|c|c|c|c|}
\hline
\textbf{Approach} & \textbf{Precision} & \textbf{Recall} & \textbf{F-Measure} \\
\hline
CodeBERT & 0.72 & 1.00 & 0.84 \\
Output Analysis & 0.88 & 0.93 & 0.90 \\ 
\textbf{Boosting} & 0.88 & 0.99 & 0.93 \\
GraphCodeBERT & 0.98 & 0.95 & 0.96 \\
\textbf{Bagging} & 0.95 & 0.97 & 0.96 \\
\hline
\end{tabular}
\caption{State-of-the-art for the IR-Plag dataset. Due to the small size of the dataset, several datasets partitions have been tested for the CodeBERT and GraphCodeBERT fine-tuning process}
\label{table:ss2}
\end{table}

Table \ref{table:ls} shows us the results of the different unsupervised similarity measures over the BigCloneBench dataset. These results estimate over 25,000 samples (6\% of the dataset) because we do not have enough computational power to analyze the complete dataset. We also discard similarity calculations in scenarios requiring more than 20 seconds of computation; otherwise, the process may require many days. Furthermore, similarity by output analysis cannot be performed in this context due to the high computational requirements of packaging, compiling, executing, and comparing several million code samples. If desired, our solution can be used on the total set, and sufficient computational power is available. 

\begin{table}
\centering
\begin{tabular}{|l|c|c|c|}
\hline
\textbf{Approach} & \textbf{Precision} & \textbf{Recall} & \textbf{F-Measure} \\ \hline
Abstract Syntax Tree & 0.15 & 0.47 & 0.23 \\
Bag-of-Words & 0.27 & 0.62 & 0.38 \\
CodeBERT & 0.18 & 0.29 & 0.23 \\
Comment Sim. & 0.20 & 0.31 & 0.25 \\
Output Analysis & 0.00 & 0.00 & 0.00 \\
Function Calls & 0.40 & 0.62 & 0.48 \\
Fuzzy Matching & 0.38 & 0.43 & 0.40 \\
Graph Matching & 0.16 & 0.46 & 0.23 \\
Rolling Hash & 0.21 & 0.23 & 0.22 \\
Perceptual Hash & 0.15 & 0.78 & 0.24 \\
Jaccard & 0.27 & 0.62 & 0.38 \\
Longest Common Subsequence & 0.22 & 0.47 & 0.30 \\
Levenshtein & 0.33 & 0.23 & 0.27 \\
Metrics comparison & 0.15 & 0.72 & 0.25 \\
N-Grams & 0.32 & 0.45 & 0.37 \\
Program Dependence Graph & 0.18 & 0.44 & 0.26 \\
Rabin-Karp & 0.15 & 0.22 & 0.18 \\
Semantic Clone & 0.23 & 0.43 & 0.30 \\
Semdiff method & 0.18 & 0.26 & 0.22 \\
TDF-IDF & 0.27 & 0.63 & 0.38 \\
Winnow & 0.20 & 0.71 & 0.31 \\ \hline
\textbf{Bagging} & 0.85 & 0.42 & 0.56 \\
\textbf{Boosting} & 0.79 & 0.45 & 0.57 \\
\hline
\end{tabular}
\caption{Summary of the results obtained for the BigCloneBench dataset}
\label{table:ls}
\end{table}

As can be seen, each technique individually shows a performance far below what would be required in a real environment. Therefore, there are better candidates for integration into real-world systems. However, their systematic aggregation through bagging or boosting techniques yields superior results, so we far exceeded our weak baseline. 

Table \ref{table:ls2} establishes a comparative for the state-of-the-art published to this dataset. For the fine-tuning of CodeBERT and GraphCodeBERT, we rely on the configuration outlined by the authors of the original papers. At the same time, the numbers relative to our strategies are estimated. The reason is that the bagging and boosting strategies have been tested using the seven most promising unsupervised similarity measures from the 415,416 samples of the test partition. These promising measures have been obtained from a preliminary study; please refer to the source code for more information. The reason is, again, not having enough computational resources to operate over the whole search space. 

\begin{table}
\centering
\begin{tabular}{|l|c|c|c|}
\hline
\textbf{Approach} & \textbf{Precision} & \textbf{Recall} & \textbf{F-Measure} \\ \hline
Deckard \cite{jiang2007deckard} & 0.93 & 0.02 & 0.03 \\
RtvNN \cite{white2016deep} & 0.95 & 0.01 & 0.01 \\
\textbf{Bagging} & 0.85 & 0.42 & 0.56 \\
\textbf{Boosting} & 0.79 & 0.45 & 0.57 \\
CDLH \cite{wei2017supervised} & 0.92 & 0.74 & 0.82 \\
ASTNN \cite{zhang2019novel} & 0.92 & 0.94 & 0.93 \\
CodeBERT \cite{key-codebert} &	0.95 &	0.93 &	0.94 \\
FA-AST-GMN \cite{wang2020detecting} & 0.96 & 0.94 & 0.95 \\
TBBCD \cite{yu2019neural} & 0.94 & 0.96 & 0.95 \\
GraphCodeBERT\cite{guo2020graphcodebert} & 0.95 &	0.95 &	0.95 \\
\hline
\end{tabular}
\caption{State-of-the-art for the BigCloneBench dataset}
\label{table:ls2}
\end{table}

\section{Discussion}
Our experiments demonstrate the effectiveness of ensemble learning for assessing code similarity. Although the results are questionable when the training data are abundant, equivalent performance is observed when working with small datasets. 

In any case, integrating diverse unsupervised similarity measures seems effective since our approach surpasses the performance of single-measure strategies regarding code clone detection. Our ensemble approach mitigates the inherent limitations of individual similarity measures. Aggregating results from multiple measures compensate for the shortcomings of any single similarity measure. For instance, a structure-focused measure may overlook semantically equivalent code with superficial syntactic differences. 

Furthermore, the adaptable nature of our ensemble approach offers other significant benefits. We can incorporate a variety of similarity measures, enabling us to tailor the approach for specific applications. If plagiarism detection is the primary objective, structure-focused measures could be given higher weight. Semantic-oriented measures might be prioritized for code reuse identification, etc. In general, our research yielded some lessons learned, detailed below.

\subsection {Lessons Learned}
Our experiments show that some unsupervised code similarity methods work reasonably well for finding source code clones. Therefore, it could make sense to integrate these methods to help improve software development under certain circumstances. For example, 

\begin{enumerate}
	\item Our ensembles can find duplicate code to rewrite as reusable parts at a low cost. This is important since code reuse is vital to many software-related tasks.

	\item Real-world code can be unstructured but our ensembles of unsupervised methods can manage this noise and variation, which is particularly relevant in scenarios with little labeled data.

	\item A human operator cannot easily interpret the Transformers-like architectures on which CodeBERT is based. Moreover, they have a huge carbon footprint when trained \cite{key-martinez-jfis}. Our approach may be reasonable in situations where these two aspects are critical.
\end{enumerate}

However, there are some limitations and areas for improvement where the performance of our ensembles is not optimal, which are discussed below.

\subsection{Limitations}
Our proposed approach has yielded promising results; however, some aspects still need improvement. First, when dealing with abundant training data, the performance of our ensembles falls substantially below the benchmarks set by state-of-the-art models.

Second, scalability presents a challenge; as codebases expand in size and complexity, the computational burden of implementing multiple similarity measures increases considerably. This escalation demands the integration of more efficient optimization techniques to manage and mitigate the increased processing requirements.

Lastly, while our ensemble method offers theoretical interpretability, its practical use demands a more pragmatic approach. Further work in this direction is crucial to demonstrate how the model identifies similar code fragments, improving transparency and usability in practice.

\section{Conclusions and Future Work}
Evaluating code similarity is a crucial component of many software development tasks. In this study, we have explored unsupervised similarity measures, leveraging their independence from labeled data and adaptability to diverse coding styles. Our findings show the efficacy of ensemble learning approaches in code similarity assessment. We have developed a strategy that surpasses traditional techniques by aggregating multiple measures, achieving performance comparable to state-of-the-art standards in the absence of significant training data.

As code becomes more complex, there is a growing need for accurate and efficient ways to measure code similarity. The software industry's ever-changing nature shows us the vital role that unsupervised similarity measures could play in this context. While supervised approaches are superior in performance, unsupervised techniques often offer advantages of another kind.

Future work must further explore the potential of unsupervised code clone detection methods. Future research could also evaluate the efficacy of other ensemble techniques and examine the applicability of transfer learning for improved performance. The ultimate goal is to advance code analysis automation. This development could significantly reduce the need for manual intervention across all phases of the software development process, thereby increasing efficiency and productivity in the field.

\section*{Acknowledgments}
This research has been funded by the Federal Ministry for Climate Action, Environment, Energy, Mobility, Innovation, and Technology (BMK), the Federal Ministry for Digital and Economic Affairs (BMDW), and the State of Upper Austria in the frame of SCCH, a center in the COMET - Competence Centers for Excellent Technologies Programme managed by Austrian Research Promotion Agency FFG.

\section*{Data availability and access}
The data and the source code to reproduce this research are available at: \url{https://github.com/jorge-martinez-gil/ensemble-codesim}. The manuscript is also available as a preprint at \url{https://arxiv.org/abs/2405.02095}.

\bibliographystyle{plain}
\bibliography{mybib}
\end{document}